\documentclass[aps,pra,twocolumn,superscriptaddress,showpacs,floatfix]{revtex4}
\usepackage{epstopdf}
\usepackage{graphicx}
\usepackage{dcolumn}
\usepackage{bm}
\usepackage{amsfonts}
\usepackage{amsmath}

\begin{document}
\renewcommand{\thefigure}{\arabic{figure}}
\def\be{\begin{equation}}
\def\ee{\end{equation}}
\def\ber{\begin{eqnarray}}
\def\eer{\end{eqnarray}}

\def\kv{{\bf k}}
\def\qv{{\bf q}}
\def\pv{{\bf p}}
\def\sigmav{{\bf \sigma}}
\def\tauv{{\bf \tau}}
\newcommand{\h}[1]{{\hat {#1}}}
\newcommand{\hdg}[1]{{\hat {#1}^\dagger}}
\newcommand{\bra}[1]{\left\langle{#1}\right|}
\newcommand{\ket}[1]{\left|{#1}\right\rangle}

\title{Theory of correlations in strongly interacting fluids of two-dimensional dipolar bosons}
\date{\today}
\author{Saeed H. Abedinpour}
\email{abedinpour@iasbs.ac.ir}
\affiliation{Department of Physics, Institute for Advanced Studies in Basic Sciences (IASBS), Zanjan 45137-66731, Iran}
\affiliation{School of Physics, Institute for Research in Fundamental Sciences (IPM), Tehran 19395-5531, Iran}
\author{Reza Asgari}
\affiliation{School of Physics, Institute for Research in Fundamental Sciences (IPM), Tehran 19395-5531, Iran}
\author{Marco Polini}
\affiliation{NEST, Istituto Nanoscienze-CNR and Scuola Normale Superiore, I-56126 Pisa, Italy}

\begin{abstract}
Ground-state properties of a two-dimensional fluid of bosons with repulsive dipole-dipole interactions are 
studied by means of the Euler-Lagrange hypernetted-chain approximation.
We present a self-consistent semi-analytical theory of the pair distribution function $g(r)$ and ground-state energy of this system. Our approach is based on the solution of a zero-energy scattering Schr\"{o}dinger equation for the ``pair amplitude'' $\sqrt{g(r)}$ with an effective potential from Jastrow-Feenberg correlations. We find excellent agreement with quantum Monte Carlo results over a wide range of coupling strength, nearly up to the critical coupling for the liquid-to-crystal quantum phase transition. We also calculate the one-body density matrix and related quantities, such as the momentum distribution function and the condensate fraction.
\end{abstract}

\maketitle

\section{Introduction}\label{sect:intro}

Although experiments on ultra-cold atomic gases are performed in an extremely dilute regime, inter-particle interactions can play an important role in determining their behavior~\cite{ref:pethick_book}. In the ultra-cold regime the van der Waals interaction between atoms can be replaced by a Fermi pseudo-potential whose sign and strength can be tuned at will by means of a Feshbach resonance~\cite{ref:pethick_book}.

Recent experimental advances~\cite{ref:dipole_bec} in cooling down atoms with permanent dipole moments and polar molecules have made it possible to study quantum gases with long-range interactions. These systems have attracted a large fraction of theoretical and experimental interest~\cite{ref:lahaye_rpp09,ref:baranov_phyrep08}.
The fact that the strength and sign of the short range part of the inter-molecular interaction can be fully controlled by means of Feshbach resonances makes polar molecules an excellent system for studying the interplay between short-range and long-range interactions. Moreover the short-range component of the scattering potential can be completely switched off, leaving us with a purely dipolar system. The dipole-dipole interaction is peculiar in the sense that it is both long range and anisotropic. The latter property is interesting in particular since the dipolar interaction can become either attractive or repulsive depending on the geometry of the external confining potential.

Magnetic dipole moments of atoms lead to an anisotropic and long-ranged
dipole-dipole interaction and experiments have shown that the
dipole-dipole interaction can influence the shape and
stability of a quantum gas~\cite{kock}. Furthermore, there has been  a lot of experimental progress 
in stabilizing and cooling dipolar molecules~\cite{ex} with an electric dipole moment
which is larger than their magnetic dipole moment, making
the dipolar interaction the dominant one. Dipole-dipole interactions between molecules can
be thus controlled by external electric fields.

A number of theoretical and computational studies have addressed ground-state 
properties of two-dimensional (2D) repulsive dipolar bosons at zero and
low temperatures~\cite{mora,ref:astrakharchik_prl07,ref:buchler}.
Quantum Monte Carlo (QMC) studies~\cite{ref:astrakharchik_prl07} at zero temperature have covered, in particular, the
whole range of coupling strength up to the crystallization transition. 

In this Article we present a theoretical study of ground-state properties of a polarized dipolar fluid of bosons with average density $n$, tightly trapped in the direction of polarization. The system is effectively 2D and the dipole-dipole interaction is isotropic. The main focus of this Article is on the so-called pair distribution function (PDF)~\cite{Giuliani_and_Vignale} $g(r)$, which is defined so that the quantity $2\pi n g(r)dr$ gives the average number of dipoles lying within a circular shell of radius $dr$ centered on a ``reference" dipole sitting at the origin. We present a self-consistent semi-analytic theory of the PDF, which incorporates many-body correlation effects, allowing us to explore the physics of the system at strong coupling. Our approach, which is based on the so-called Euler-Lagrange hypernetted-chain (HNC) approximation~\cite{lantto,zab,report}, involves the solution of a zero-energy scattering Schr\"{o}dinger equation with an effective potential~\cite{ref:kallio, davoudi_prb_2003_fermions, davoudi_prb_2003_bosons,ref:asgari_ssc2004,abedinpour_ssc_2007}. Following Ref.~\cite{ref:asgari_ssc2004}, we transcend the HNC approximation by taking into account higher-order ({\it i.e.} three-body) correlations in an effective way. We also calculate,  within the lowest-order HNC formalism,  the one-body density matrix, the momentum distribution function, and the condensation fraction.

A similar study has recently been carried out by Hufnagl {\it et al.}~\cite{ref:hufnagl_jltp10}. The authors of this work have taken into account both the anisotropy of the dipolar interaction and the short-range part of the repulsive potential, but have neglected three-body correlations, thereby obtaining very good agreement with QMC data only in the weak-coupling regime.

The outline of this paper is as follows. In Sec.~\ref{sect:dipolar} we briefly review the main properties of the dipole-dipole interaction.
In Sec.~\ref{sect:theory} we present a formally exact zero-energy scattering equation for the pair amplitude $\sqrt{g(r)}$ and introduce the approximations that we employ for the evaluation of the effective potential. Our main numerical results are presented in Sec.~\ref{sect:num}. Finally, Sec.~\ref{sect:concl} summarizes our main conclusions.

\section{Dipolar interactions}
\label{sect:dipolar}

The interaction energy between two particles with identical dipole moments aligned 
along the unit vectors ${\bm n}_1$ and ${\bm n}_2$ and placed at a distance ${\bm r}_{12}$ from each other is given by the following well-known expression:
\be\label{eq:udd_gen}
v_{\rm dd}({\bm r}_{12})=\frac{C_{\rm dd}}{4\pi}~\frac{({\bm n}_1\cdot {\bm n}_2 ) r^2_{12} - 3 ({\bm n}_1 \cdot {\bm r}_{12})({\bm n}_2\cdot {\bm r}_{12})}{r^5_{12}}~.
\ee
Here $C_{\rm dd}$ is the dipole-dipole coupling constant, which depends on the microscopic origin of the interaction: {\it e.g.}, it is $d^2/\epsilon_0$ for particles with permanent electric dipole $d$ and $\mu_0 M^2$ for particles with permanent magnetic dipole $M$ (here $\epsilon_0$ and $\mu_0$ are the permittivity and permeability of vacuum, respectively). An electron-hole bilayer in the exciton condensate phase~\cite{doublelayerEC} is another realization of a system with dipolar interactions. In this case $C_{\rm dd} = e^2 d^2/\epsilon$, where $-e$ is the electron's charge, $d$ is the inter-layer separation, and $\epsilon$ is the dielectric constant of the host semiconductor.

For the polarized case, where all dipoles align in the same direction, Eq.~(\ref{eq:udd_gen}) simplifies to
\be\label{eq:udd_pol}
v_{\rm dd}({\bm r}_{12}) = \frac{C_{\rm dd}}{4\pi}~\frac{1-3 \cos^2(\theta)}{r^3_{12}}~.
\ee
This interaction has two important features. It is long-ranged, {\it i.e.} it decays like $1/r^3$ at large distances, and furthermore, it is anisotropic. In particular, it is repulsive for dipoles aligned side-by-side ($\theta=\pi/2$) and is attractive for dipoles aligned head-to-toe ($\theta=0$). 
In this work we study 2D systems of polarized dipoles, so that the inter-particle interaction is isotropic:
\be\label{eq:udd_2d}
v_{\rm dd}(r_{12}) = \frac{C_{\rm dd}}{4\pi}~\frac{1}{r^3_{12}}~.
\ee
\section{Theory}
\label{sect:theory}

We consider a 2D fluid of $N$ point dipoles with {\it bosonic} statistics. The first-quantized Hamiltonian can be written as~\cite{ref:astrakharchik_prl07}:
\begin{equation}\label{eq:hamil}
{\cal H} =-\frac{\hbar^2}{2m}\sum_i\nabla^{2}_{{\bm r}_i} + \frac{C_{\rm dd}}{4\pi}\sum_{i<j}\frac{1}{|{\bm r}_{i} - {\bm r}_{j}|^3}~,
\end{equation}
$m$ being the mass of a nano-particle dipole. The ground-state properties of the Hamiltonian~(\ref{eq:hamil}) are governed by a single dimensionless parameter:
\be\label{eq:couplingconstant}
\gamma=n r^2_0~,
\ee
where $n$ is the average density and $r_0=m C_{\rm dd}/(4\pi\hbar^2)$ is a characteristic length scale, which is typically of the order of a few Angstroms.

In order to calculate the ground-state properties of the Hamiltonian (\ref{eq:hamil}), we use the HNC~\cite{lantto,zab,report} approximation 
at zero temperature. In what follows we first present our theory at the simplest level (which works well in the limit $\gamma \ll 1$) and
then transcend it to obtain accurate results at strong coupling ($\gamma \gg 1$).

With the zero of energy taken at the chemical potential, the
formally exact differential equation for the pair-correlation
function can be written as~\cite{davoudi_prb_2003_fermions,davoudi_prb_2003_bosons}
\begin{equation}\label{eq:scat}
\left[-\frac{\hbar^2}{m}\nabla^2_{\bm r} + V_{\rm eff}(r)\right]\sqrt{g(r)}=0~.
\end{equation}
The effective scattering potential is 
\be\label{eq:effectivepot}
V_{\rm eff}(r)=v_{\rm dd}(r)+W_{\rm B}(r)~,
\ee
where $v_{\rm dd}(r)$ is the bare repulsive dipole-dipole interaction in Eq.~(\ref{eq:udd_2d}) and $W_{\rm B}(r)$ is defined, at the level of the so called ``HNC/0" approximation, 
by the following equation~\cite{ref:chakraborty}:
\be\label{eq:wb}
W_{\rm B}(k) = -\frac{\varepsilon(k)}{2}\left[2S(k)+1\right]\left[\frac{S(k)-1}{S(k)}\right]^2~.
\ee
In writing Eq.~(\ref{eq:wb}) we have introduced the Fourier transform (FT) $W_{\rm B}(k)$ of $W_{\rm B}(r)$ according to 
\be
{\rm FT}[F(r)] \equiv n\int d^2{\bm r}~F(r) \exp{(i{\bm k}\cdot {\bm r})}~.
\ee 
Furthermore, $\varepsilon(k)=\hbar^2k^2/(2m)$ is the single-particle energy and $S(k)$ is the instantaneous or ``static" structure factor~\cite{Giuliani_and_Vignale}, $S(k)=1+ {\rm FT}[g(r)-1]$. Eqs.~(\ref{eq:scat})-(\ref{eq:wb}) form a closed set of equations, which can be solved numerically self-consistently to the desired degree of accuracy.

When $\gamma \sim 1$ the simplest approximation for $W_{\rm B}(r)$ in Eq.~(\ref{eq:wb}) is inadequate. 
Improvements on Eq.~(\ref{eq:wb}) for a Bose fluid can be sought in two directions~\cite{ref:apaja_w3}. The HNC/0 may be transcended by the inclusion of i) low-order ``elementary" diagrams and ii) three-body Jastrow-Feenberg correlations. 

The contribution from three-body correlations is~\cite{ref:apaja_w3}
\ber\label{eq:w3}
W_{\rm B}^{(3)}(k) &=& \frac{1}{4n}\int \frac{d^2{\bm q}}{(2\pi)^2}~S(p)S(q)u_{3}({\bm q}, {\bm p},{\bm k})\big\{\nu_{3}({\bm q}, {\bm p}, {\bm k}) \nonumber\\
&+& [E(p)+E(q)]u_{3}({\bm q}, {\bm p}, {\bm k})\big\}~.
\eer
In Eq.~(\ref{eq:w3}) ${\bm p} = -({\bm q} + {\bm k})$, $E(k)=\varepsilon(k)/S(k)$ is the Feynman-Bijl excitation spectrum (see Sect.~\ref{sect:num}), 
\be\label{eq:nu3}
\nu_{3}({\bm q}, {\bm p}, {\bm k}) = (\hbar^2/m)[{\bm k}\cdot{\bm p}\chi(p) + {\bm k}\cdot{\bm q}\chi(q)+{\bm p}\cdot{\bm q}\chi(q)]~,
\ee
and
\ber\label{eq:u3}
u_{3}({\bm q}, {\bm p}, {\bm k}) &=& -\frac{(\hbar^2/2m)}{E(k)+E(p)+E(q)}\nonumber\\
&\times& [{\bm k}\cdot{\bm p}\chi(k)\chi(p) +{\bm p}\cdot{\bm q}\chi(p)\chi(q) \nonumber \\ 
&+ & {\bm k}\cdot{\bm q}\chi(k)\chi(q)]~.
\eer
In Eqs.~(\ref{eq:nu3})-(\ref{eq:u3}) $\chi(k)=1-1/S(k)$.

We have taken into account higher-order terms that are missed by the HNC/0 approximation by assuming that they lead to corrections to the scattering potential $V_{\rm eff}(r)$. Using the theory developed by Apaja {\it et al.}~\cite{ref:apaja_w3} we have supplemented $W_{\rm B}(k)$ in Eq.~(\ref{eq:wb}) by the inclusion of the three-body potential $W^{(3)}_{\rm B}(k)$:
\be\label{eq:correlation_w3}
W_{\rm B}(k) \to W_{\rm B}(k)+ \alpha(\gamma)W_{\rm B}^{(3)}(k)~.
\ee
If $\alpha(\gamma)$ is set to unity, the r.h.s. of Eq.~(\ref{eq:correlation_w3}) defines the so-called ``HNC/$3$" approximation. It has been shown~\cite{ref:asgari_ssc2004} that higher-order corrections beyond HNC/$3$ can be effectively taken into account by introducing a weighting function $\alpha(\gamma) > 1$. This approximation will be termed ``HNC/$\alpha3$".  

The functional dependence of the weighting factor $\alpha(\gamma)$ on the coupling constant $\gamma$ can be fixed, for example~\cite{ref:asgari_ssc2004}, by requiring that the ground-state energy per particle extracted from the PDF calculated within the HNC/$\alpha3$ approximation matches exactly the corresponding quantity calculated by QMC simulations~\cite{ref:astrakharchik_prl07}. Following this procedure we have calculated numerically the function $\alpha(\gamma)$ and produced a convenient analytical parametrization of it in the interval $1 \leq \gamma \leq 256$: 
\be\label{eq:parametrization}
\alpha(\gamma) = 1.88+3.26\exp{(-0.26 \gamma^{0.56})}~.
\ee 

Numerical results for the PDF calculated within the three approximations described in this Section (HNC/$0$, HNC/$3$, and HNC/$\alpha3$) will be illustrated in Sect.~\ref{sect:num} and severely tested against accurate QMC data by Astrakharchik {\it et al.}~\cite{ref:astrakharchik_prl07}. Practical recipes on how to solve Eq.~(\ref{eq:scat}) are discussed in detail in Ref.~\cite{davoudi_prb_2003_fermions}.

\begin{figure}
\includegraphics[width=1.0\linewidth]{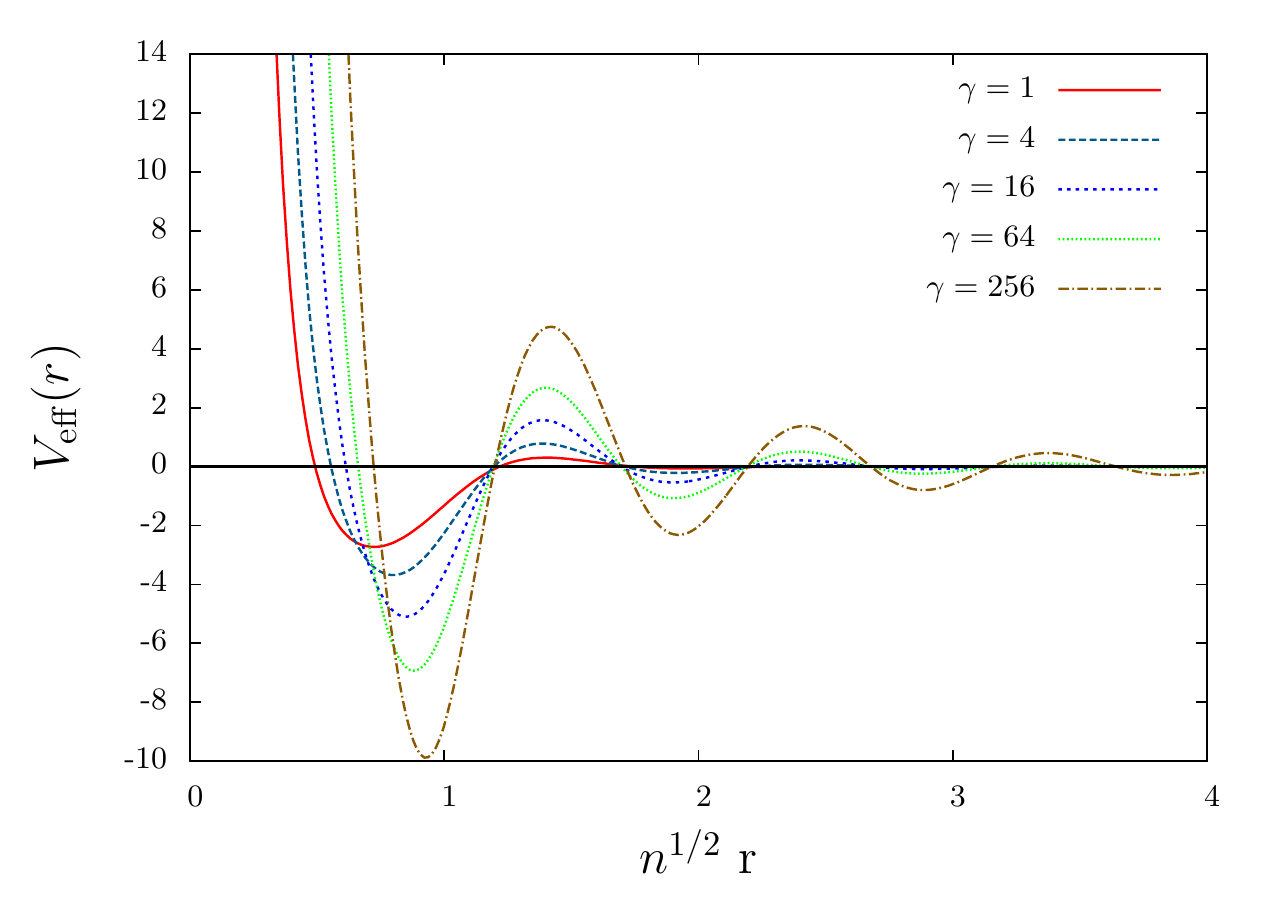}
\caption{(Color online) The effective scattering potential $V_{\rm eff}(r)$ of a 2D fluid of dipolar bosons [in units of $\gamma\hbar^2 /(m r_0^2)$] as a function of $n^{1/2} r$. Different curves correspond to different values of the dimensionless coupling constant $\gamma$. All the curves in this plot have been calculated within the HNC/$\alpha 3$ approximation.\label{fig:veff}}
\end{figure}
\begin{figure*}
\tabcolsep=0 cm
\begin{tabular}{cc}
\includegraphics[width=0.5\linewidth]{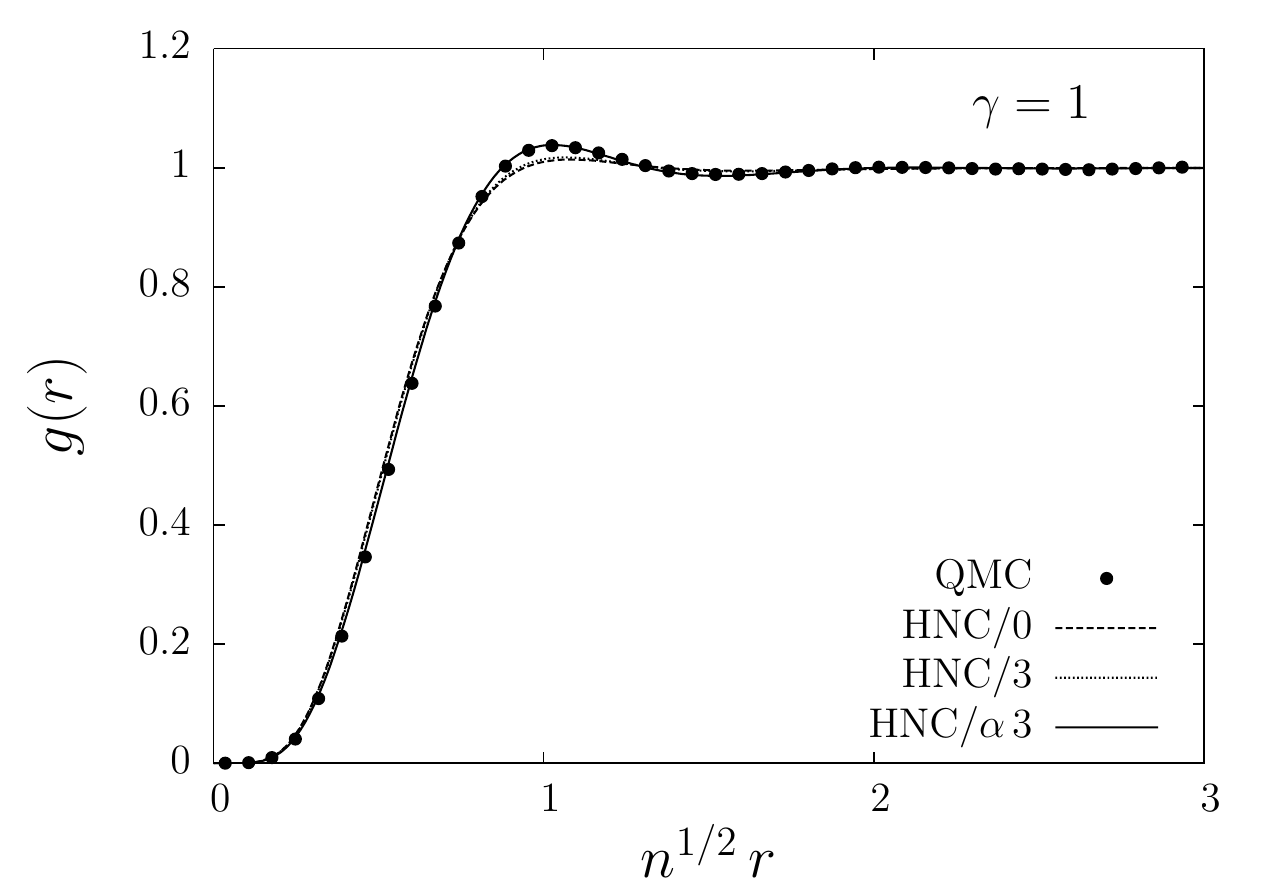}&
\includegraphics[width=0.5\linewidth]{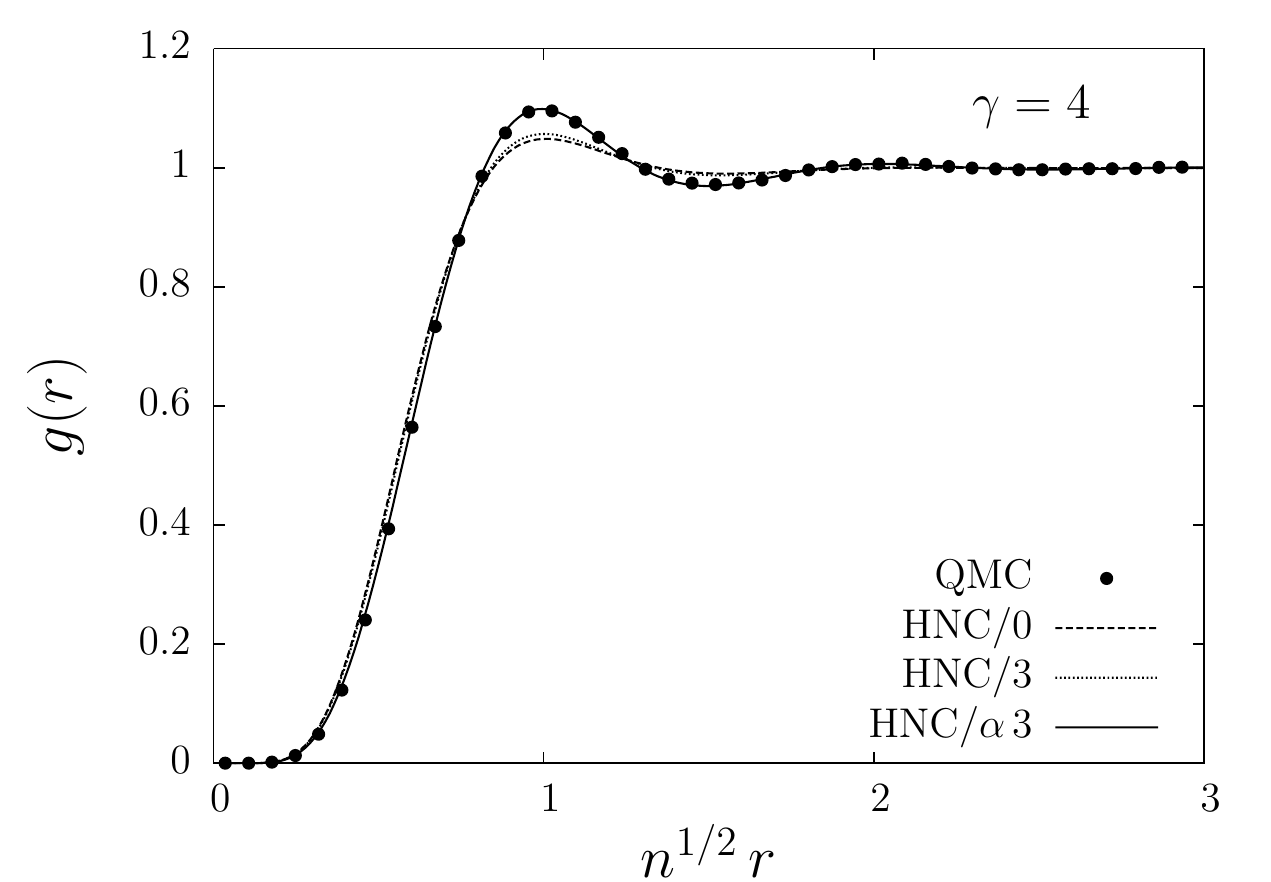}\\
\includegraphics[width=0.5\linewidth]{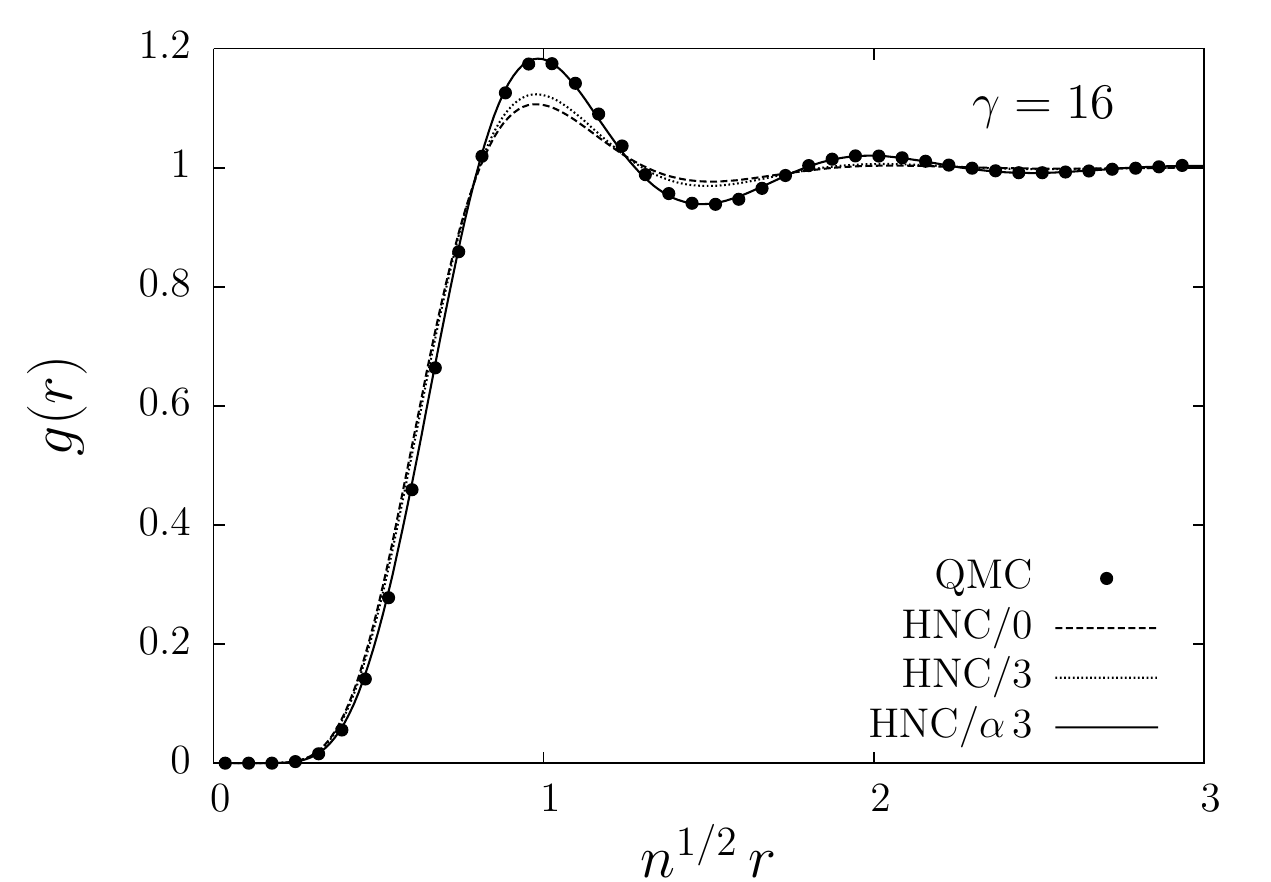}&
\includegraphics[width=0.5\linewidth]{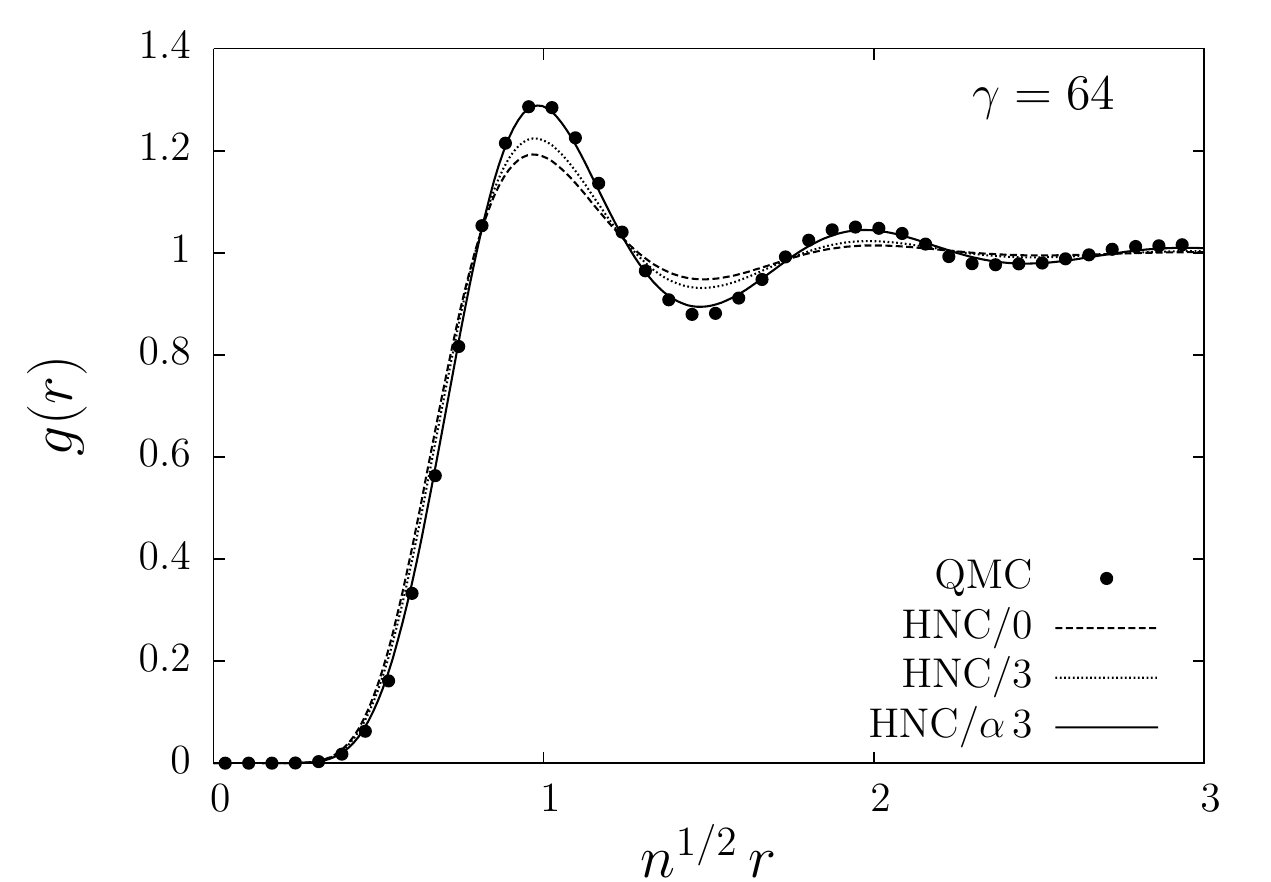}\\
\includegraphics[width=0.5\linewidth]{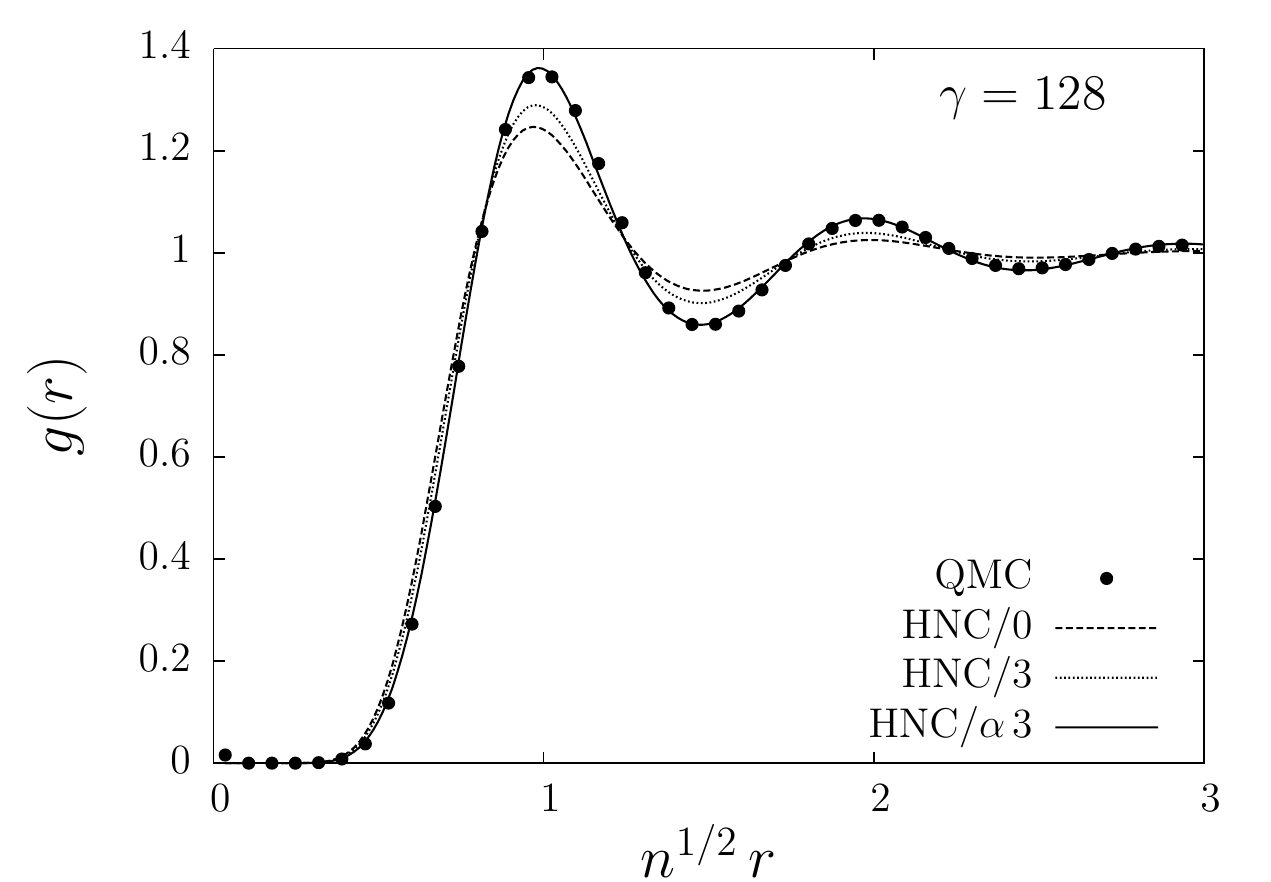}&
\includegraphics[width=0.5\linewidth]{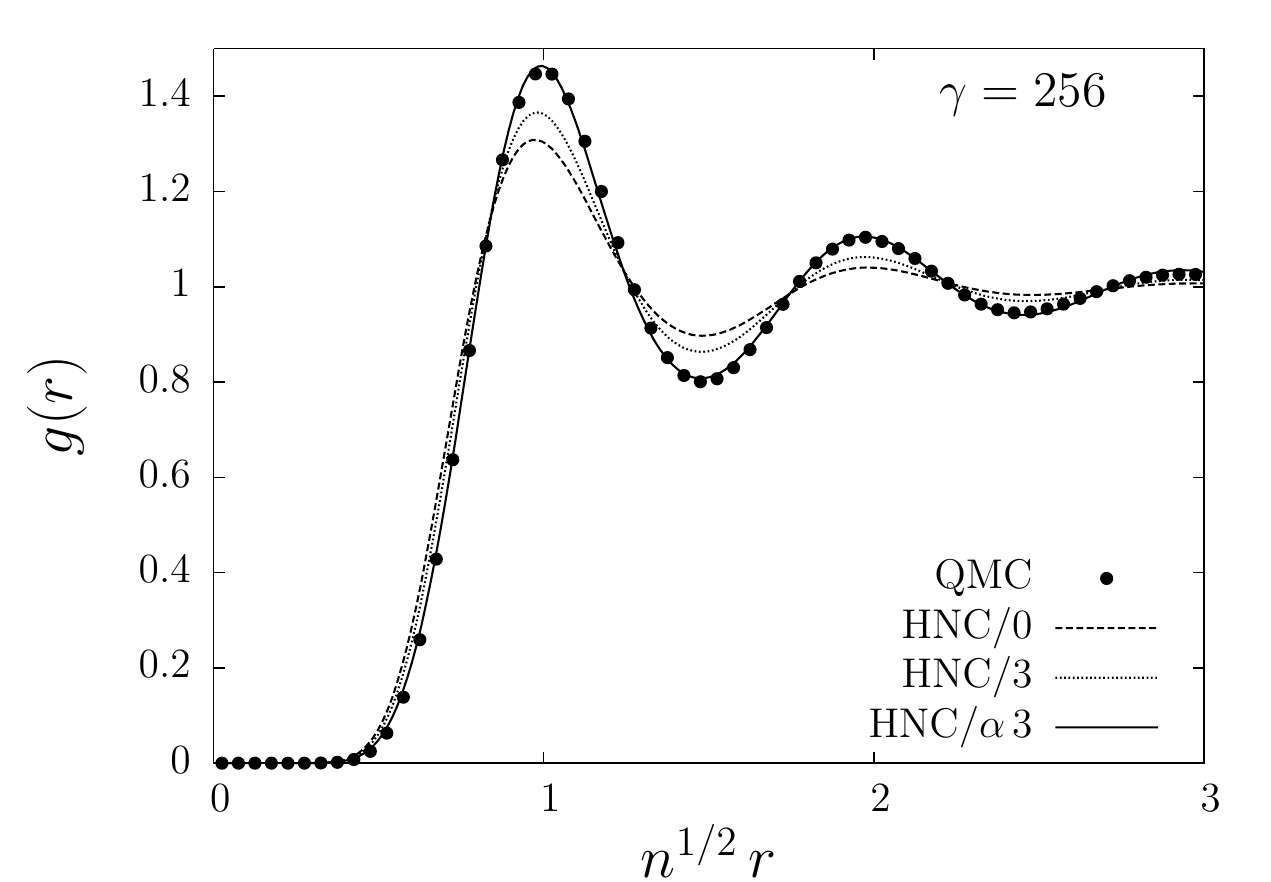}
\end{tabular}
\caption{The pair distribution function $g(r)$ of a 2D fluid of dipolar bosons as a function of $n^{1/2}r$ for different values of $\gamma$. Dashed, dotted, and solid lines represent theoretical results based on the HNC/$0$, HNC/$3$, and HNC/$\alpha3$ approximations, respectively. Filled circles denote QMC results from Ref.~\cite{ref:astrakharchik_prl07}.\label{fig:gr}}
\end{figure*}
\subsection{The one-body density matrix}
\label{sect:momentum}

Although the main focus of this Article is on the PDF, in this Section we briefly discuss how to calculate the one-body density matrix (1BDM) 
and other interesting quantities on the basis of the simplest scheme discussed earlier, {\it i.e.} the HNC/$0$. 

Once the PDF $g(r)$ is obtained from the formalism outlined in the previous Section, one can use it to find the 1BDM and hence the condensate fraction and the momentum distribution function. 
The 1BDM of a homogenous quantum liquid is related to its many-body wave function $\Psi({\bm r}_1, {\bm r}_2,\dots, {\bm r}_N)$ by the following relation:
\ber\label{eq:1bdm}
\rho(r) &=&
\int d^2{\bm r}_2 \dots d^2{\bm r}_N \Psi^\star({\bm r}, {\bm r}_2,\dots, {\bm r}_N)\nonumber\\
&\times& \Psi({\bm 0}, {\bm r}_2,\dots, {\bm r}_N)~,
\eer
where we have assumed the following normalization $\int d^2{\bm r}_1 \dots d^2{\bm r}_N |\Psi({\bm r}_1, {\bm r}_2,\dots, {\bm r}_N)|^2 = N$, $N$ being the total number of particles.

In a homogeneous system the value at the origin of the 1BDM gives the particle density, $\rho(0) = n$, 
while its long-distance behavior gives the condensate fraction $n_0$ ({\it i.e.} the fraction of particles occupying macroscopically the zero-momentum state, here measured in units of the total density $n$), 
$n_0 =\rho(r \to \infty)/n$. Furthermore, the momentum distribution function $n(k)$ is also related to the 1BDM via Fourier transformation:
\be\label{eq:n_k}
n(k) = n n_0(2\pi)^2\delta({\bm k})+ {\rm FT}[\rho(r)/n -n_0]~.
\ee

Within the HNC/0 formalism, the 1BDM reads~\cite{ref:fantoni_nuovo78,ref:manousakis_prb85}
\be\label{eq:rho_r}
\rho(r)=n n_0 e^{N_{\rm ww}(r)}~,
\ee
where the $N_{\rm ww}(r)$ nodal function is given (in Fourier transform) by
\be
N_{\rm ww}(k)=\left[S_{\rm wd}(k)-1\right]\left[S_{\rm wd}(k)-1-N_{\rm wd}(k)\right]~.
\ee
Here $S_{\rm wd}(k)$ and $N_{\rm wd}(k)$ could be obtained from the solution of the following coupled equations
\be\label{eq:nwd}
N_{\rm wd}(k)=\left[S_{\rm wd}(k)-1\right]\left[S(k)-1-N(k)\right]~,
\ee
\be\label{eq:swd}
S_{\rm wd}(k)=1+{\rm FT}[g_{\rm wd}(r)-1]~,
\ee
and
\be\label{eq:gwd}
g_{\rm wd}(r)=f(r) e^{N_{\rm wd}(r)}~.
\ee
Here
\be
N(k)=\frac{\left[S(k)-1\right]^2}{S(k)}
\ee
is the conventional nodal function and the correlation function $f(r)$ is given by $f(r) = g(r)\exp{[-N(r)]}$.

Once self-consistency in the solution of Eqs.~(\ref{eq:nwd})-(\ref{eq:gwd}) is achieved, we can find the condensation fraction via
\be
n_0 = \exp{(2R_{\rm w} - R_{\rm d})}~,
\ee
where
\ber
R_{\rm w}&=& n \int d^2{\bm r} \left[g_{\rm wd}(r) - 1 - N_{\rm wd}(r)\right]\nonumber \\
&-&\frac{n}{2} \int d^2{\bm r} \left[g_{\rm wd}(r)-1\right]N_{\rm wd}(r)~,
\eer
and
\ber
R_{\rm d} &=& n \int d^2{\bm r} \left[g(r)-1-N(r)\right] \nonumber \\
&-&\frac{n}{2} \int d^2{\bm r} \left[g(r)-1\right]N(r)~.
\eer
All the necessary ingredients to find the 1BDM and the momentum distribution function are now available.

\section{Numerical Results}
\label{sect:num}

We now turn to a presentation of our main numerical results. 

Fig.~\ref{fig:veff} illustrates the dependence of the effective scattering potential $V_{\rm eff}(r)$ calculated at the HNC/$\alpha3$ level on the interaction strength $\gamma$. Note that at strong coupling the behavior of $V_{\rm eff}(r)$ is highly oscillatory indicating the emergence of short-range order with increasing $\gamma$ and the incipient quantum phase transition to an ordered crystalline phase at strong enough coupling. According to the QMC study by Astrakharchik {\it et al.}~\cite{ref:astrakharchik_prl07}, $\gamma \simeq 290$ is the critical value at which this quantum phase transition occurs.

The degree of agreement between our theoretical results for the PDF and corresponding QMC data is illustrated in Fig.~\ref{fig:gr}. The inclusion of effectively-enhanced three-body correlations (HNC/$\alpha3$ theory) yields an excellent agreement with QMC data over the entire range of coupling constants where the fluid phase is stable. Note that when
two point dipoles are in close proximity to each other, the PDF is determined by the solution of a two-particle Schr\"{o}dinger equation in the $r\to 0$ limit. According to 
Kimball's cusp condition~\cite{kimball} one indeed finds $g(r\to 0)\sim \sqrt{r/r_0}\exp(-4\sqrt{r_0/r})$.

\begin{figure}
\includegraphics[width=0.9\linewidth]{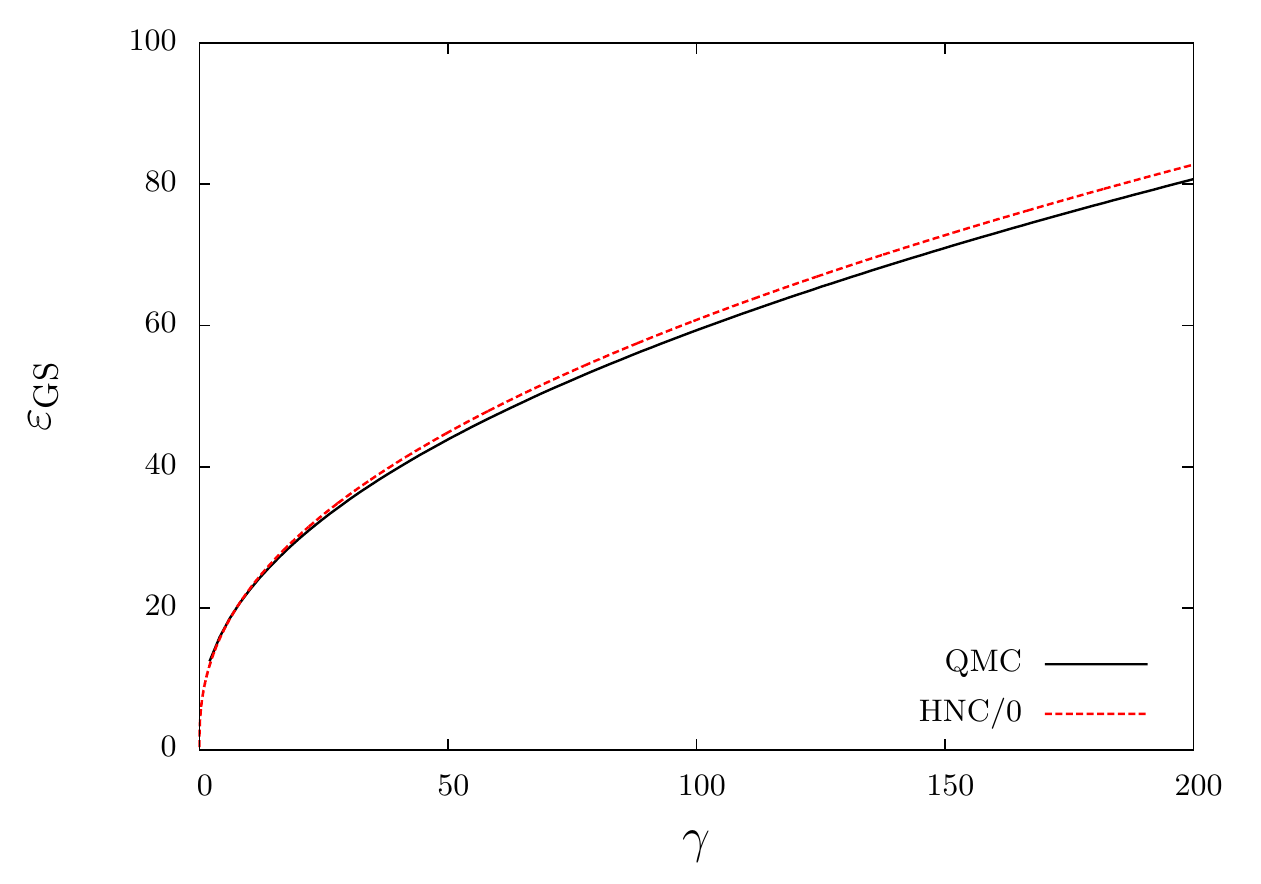}
\caption{ (Color online) The ground-state energy (per particle) of a 2D fluid of dipolar bosons $\varepsilon_{\rm GS}$ in units of $\gamma\hbar^2 /(m r_0^2)$ as a function of interaction strength $\gamma$. The dashed line represents our HNC/$0$ results. The solid curve is the parametrization formula of the QMC data reported by Astrakharchik {\it et al.}~\cite{ref:astrakharchik_prl07}. We have not plotted our HNC/$\alpha3$ results since they coincide by construction with the QMC results (we remind the reader that enforcing this constraint allows us to find the weighting factor $\alpha(\gamma)$ that enhances triplet correlations at strong coupling).\label{fig:energy}}
\end{figure}

The ground-state energy per particle $\varepsilon_{\rm GS}$ of the system can be calculated by means of the well-known integration-over-the-coupling-constant algorithm~\cite{Giuliani_and_Vignale}:
\ber
\varepsilon_{\rm GS}&=&\frac{n}{2}\int_{0}^{1}d\lambda\int d^2{\bm r}~v_{\rm dd}(r) g_{\lambda}(r)\nonumber\\
&=& \frac{nC_{\rm dd}}{4}\int_{0}^{1}d\lambda\int_{0}^{\infty} d r \frac{g_{\lambda}(r)}{r^2}~,
\eer
where $g_{\lambda}(r)$ is the PDF of an auxiliary system with dipole-dipole interactions of the form $v^{(\lambda)}_{\rm dd}(r) = \lambda v_{\rm dd}(r) = \lambda C_{\rm dd}/(4\pi r^3)$. In practice the integration over $\lambda$ is carried out by integrating over $\sqrt{\gamma}$. In Fig.~\ref{fig:energy} we have reported the ground-state energy calculated at the HNC/$0$ level in comparison with the QMC result~\cite{ref:astrakharchik_prl07}. (By construction, the ground-state energy calculated at the HNC/$\alpha3$ level coincides with the QMC result.) 

\begin{figure}
\includegraphics[width=0.9\linewidth]{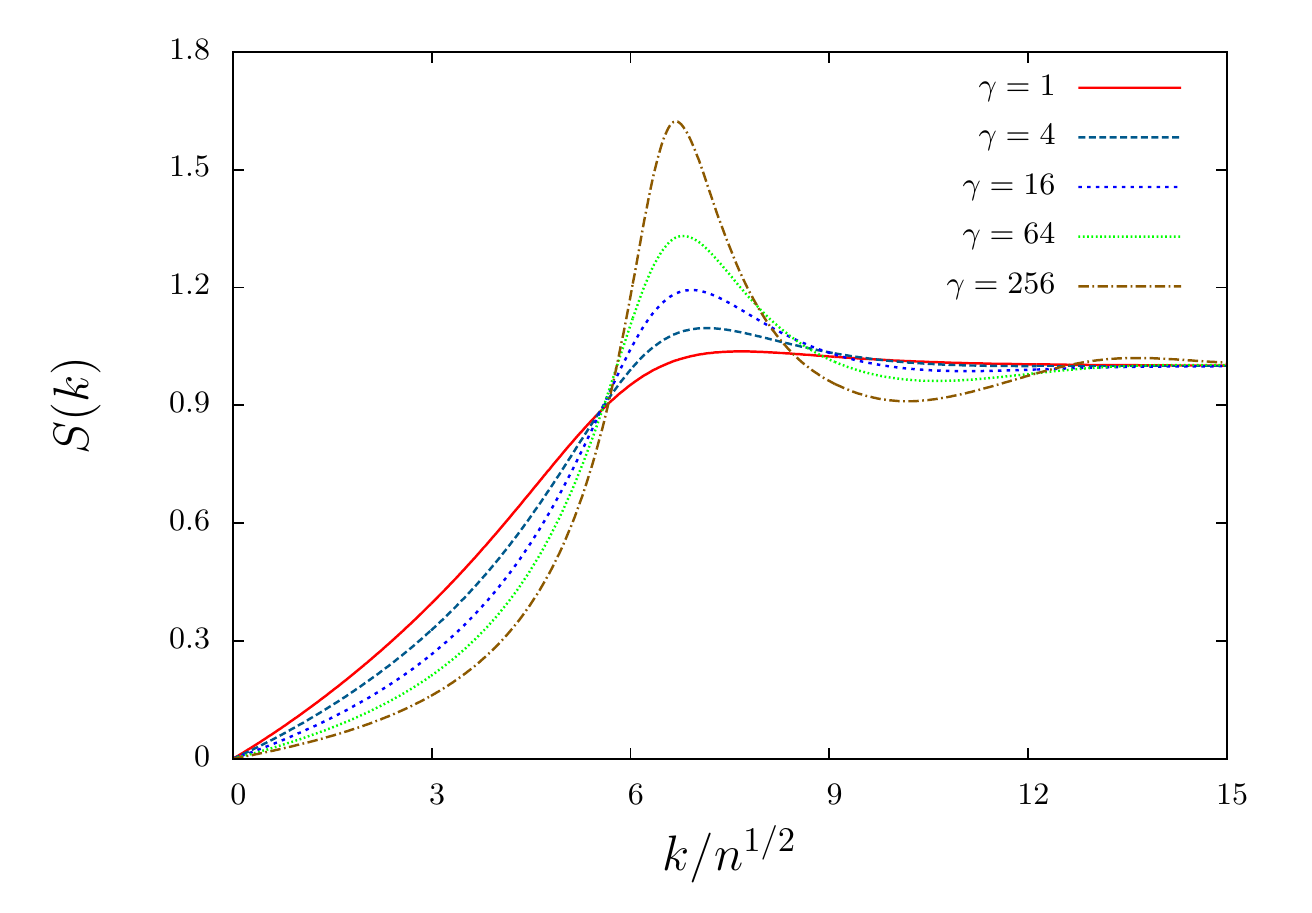}
\caption{(Color online) The instantaneous structure factor $S(k)$ of a 2D fluid of dipolar bosons as a function of $k/n^{1/2}$ for various values of $\gamma$. Data in this plot refer to the HNC/$\alpha3$ approximation.\label{fig:sp}}
\end{figure}

\begin{figure}
\includegraphics[width=0.95\linewidth]{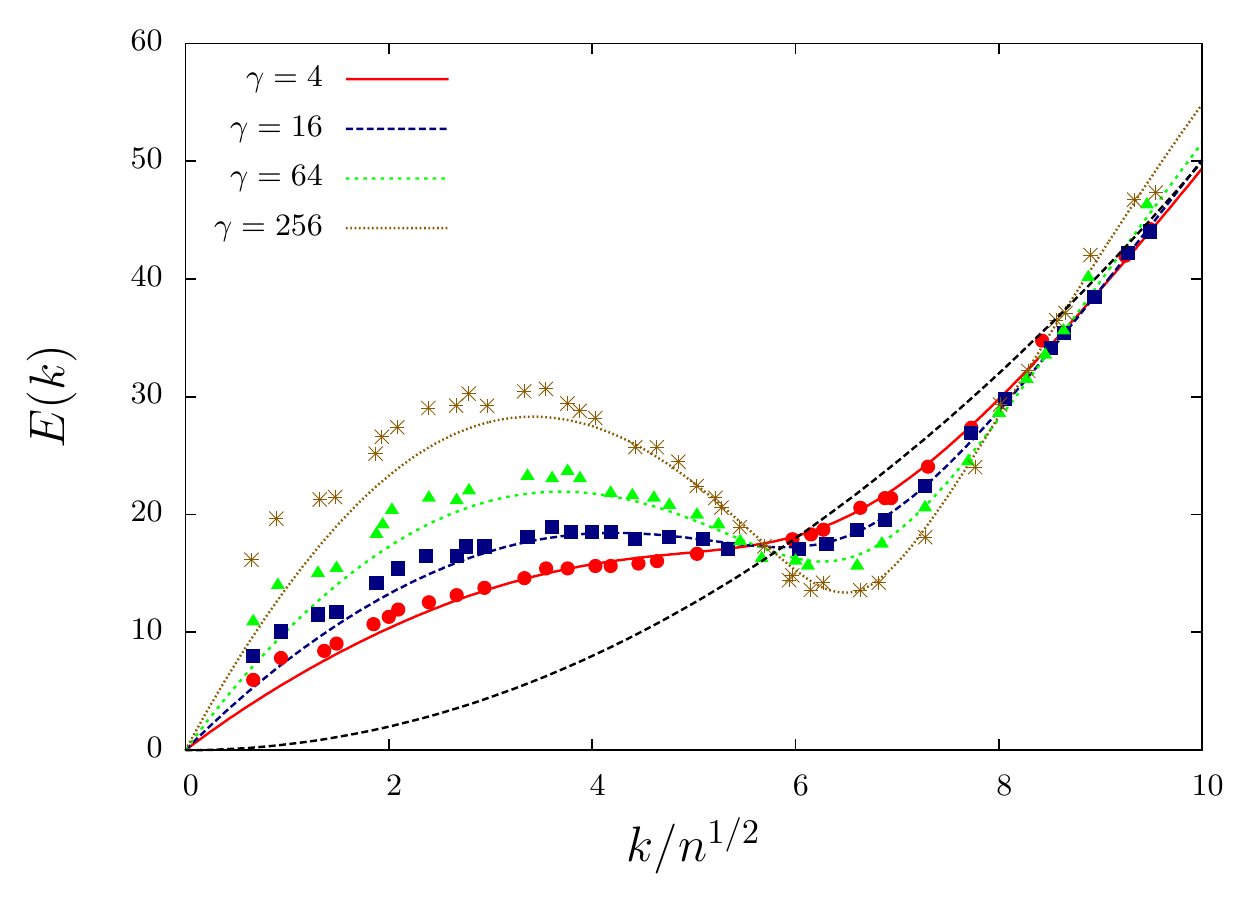}
\caption{(Color online) The upper bound $E(k)$ on the excitation spectrum of a 2D fluid of dipolar bosons in units of $\gamma\hbar^2 /(m r_0^2)$ is plotted as a function of $k/n^{1/2}$ for various values of 
$\gamma$. The (black) dashed line is the noninteracting parabolic spectrum $\varepsilon(k) = \hbar^2k^2/(2m)$. All the other lines refer to the HNC/$\alpha3$ approximation. 
Symbols indicate the prediction for $E(k)$ obtained from the QMC results~\cite{ref:astrakharchik_prl07} for $S(k)$.\label{fig:ep}}
\end{figure}

Figs.~\ref{fig:sp} and~\ref{fig:ep} illustrate the static structure factor $S(k)$ and the (approximate) excitation spectrum $E(k)$ for several values of $\gamma$, respectively.
As the coupling constant is increased, correlations get stronger and the height of the first-neighbor peak in $S(k)$ increases. Note that $S(k)$
vanishes linearly in the long-wavelength limit.

We calculate the excitation spectrum $E(k)$ from 
the approximate Feynman-Bijl relation~\cite{BF}
\begin{equation}\label{eq:ep}
E(k)=\frac{\varepsilon(k)}{S(k)}~,
\end{equation}
which is believed to be asymptotically exact for $k\to 0$ and provides a rigorous upper bound on the excitation energy at finite $k$ (since it represents the average energy of the excitations which couple to the ground state through the density and thus necessarily exceeds the minimum excitation energy~\cite{GMP}). Since $S(k\to 0) \propto k$, one finds collective excitations with an acoustic dispersion at small $k$. In agreement with the QMC data~\cite{ref:astrakharchik_prl07}, we observe that deviations from the linear phononic behavior at small $k$ start very soon and that a strong roton minimum appears at finite $k$ as the interaction strength increases. The corresponding roton gap decreases upon increasing the interaction strength. The structure of the phonon-roton peak of 2D dipolar bosons has been recently studied by Mazzanti {\it et al.}~\cite{mazzanti} by employing the dynamical structure factor $S(k,\omega)$. Moreover, multi-particle excitations, which are absent in the Feynman-Bijl theory, have also been taken into account by the same authors~\cite{mazzanti}.

\begin{figure}
\includegraphics[width=0.95\linewidth]{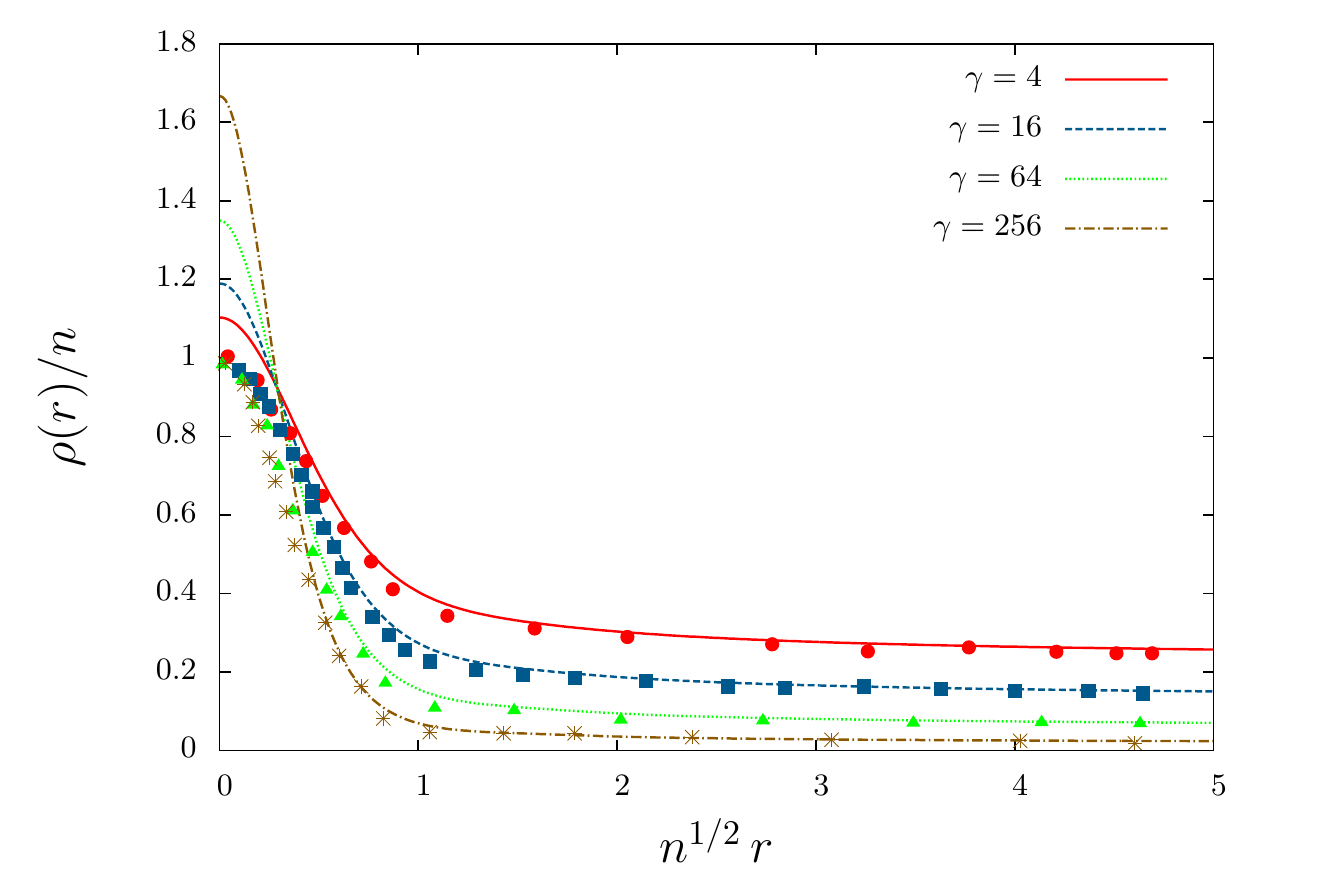}
\caption{(Color online) The one-body density matrix $\rho(r)$ of a 2D fluid of dipolar bosons (in units of $n$) as a function of $n^{1/2} r$ for various values of $\gamma$. All the theoretical results in this plot (lines) refer to the HNC/$0$ approximation. Symbols refer to QMC results from Ref.~\cite{astrakharchik_RPMBT}.\label{fig:rho_r}}
\end{figure}

\begin{figure}
\includegraphics[width=0.95\linewidth]{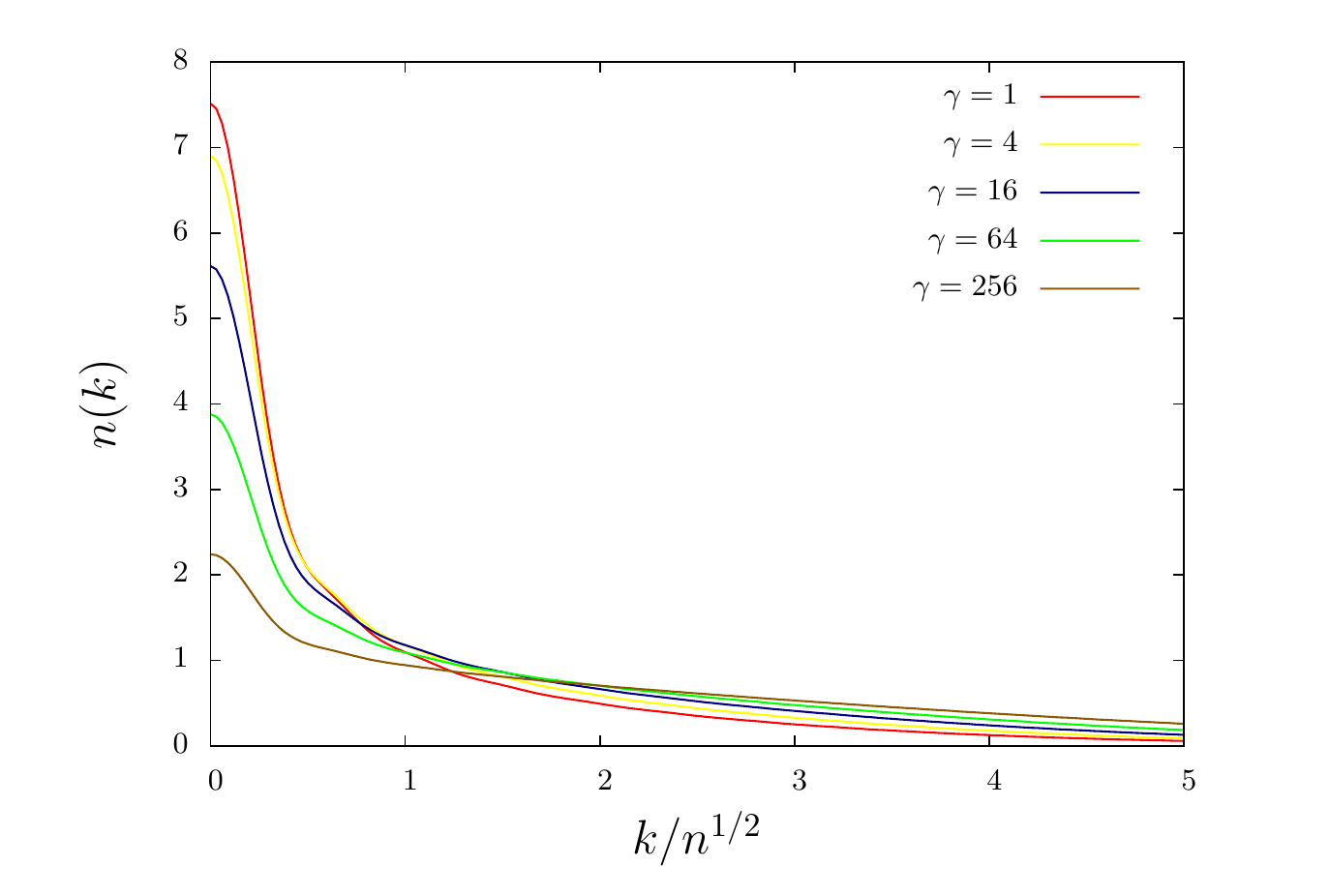}
\caption{(Color online) The momentum distribution function $n(k)$ of a 2D fluid of dipolar bosons as a function of $k/n^{1/2}$ for various values of $\gamma$. \label{fig:n_k}}
\end{figure}

\begin{figure}
\includegraphics[width=0.95\linewidth]{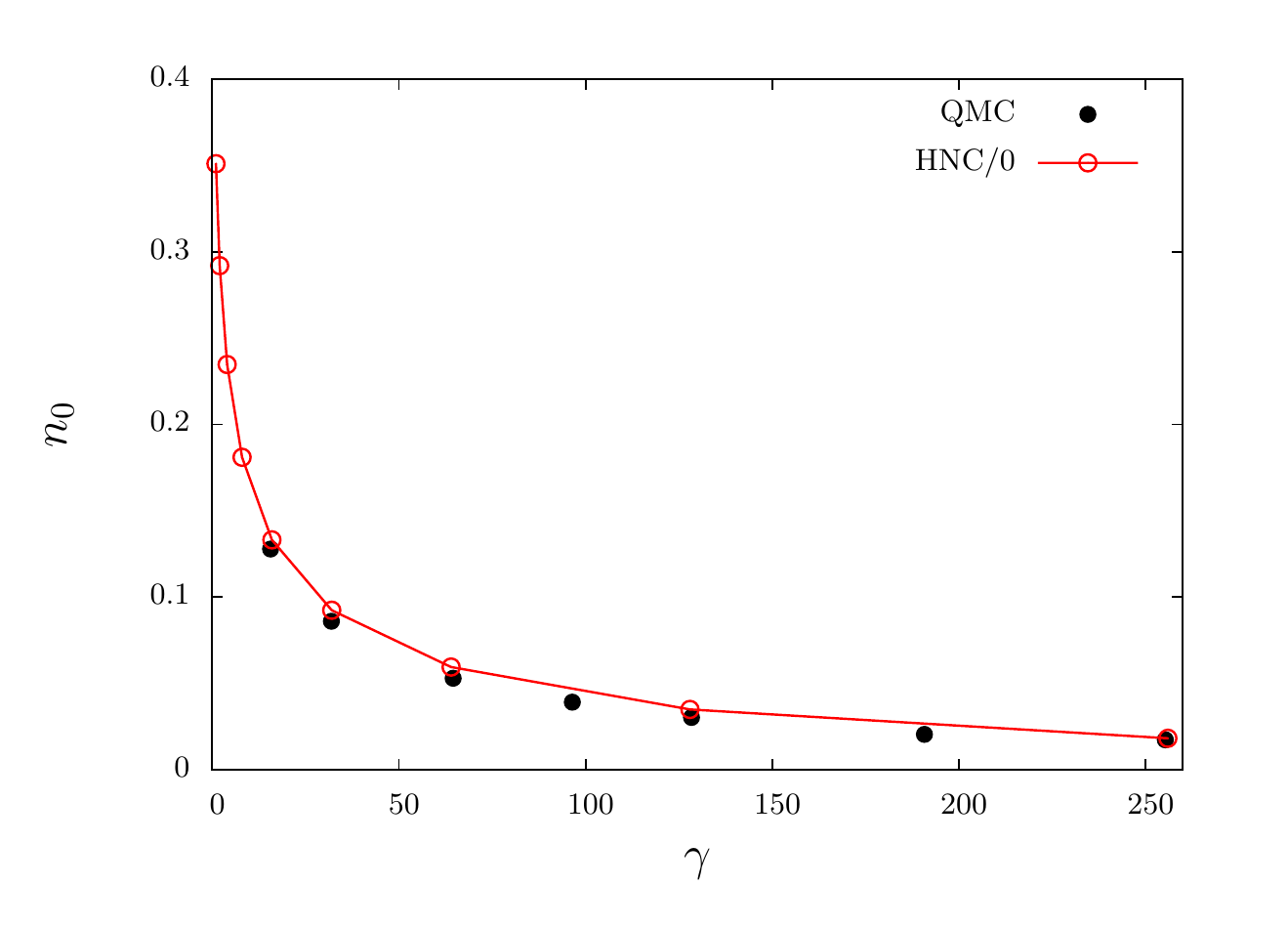}
\caption{(Color online) The condensate fraction $n_0$ of a 2D fluid of dipolar bosons as a function of $\gamma$.
 Empty circles represent the HNC/$0$ results (the line connecting the empty circles is just a guide to the eye). Filled circles are QMC data from Ref.~\cite{ref:astrakharchik_prl07}. \label{fig:n0}}
\end{figure}

Finally, in Figs.~\ref{fig:rho_r}-\ref{fig:n0} we illustrate HNC/$0$ results for the 1BDM $\rho(r)$, 
the momentum distribution function $n(k)$, and the condensate fraction $n_0$, respectively. 

Our HNC/$0$ results for the 1BDM are in good agreement with the corresponding QMC data~\cite{astrakharchik_RPMBT} at large $r$ (see Fig.~\ref{fig:rho_r}), but deviate considerably from them at short distances. Specifically, we do not recover the exact result $\rho(0)= n$. This is a well known deficiency of the HNC/$0$ approximation~\cite{ref:manousakis_prb85}.

The condensate fraction of 2D fluids of bosons is an intriguing quantity. It is well known~\cite{magro} that 2D bosons with $\ln(r)$ interactions, despite displaying superfluidity, are characterized by a vanishing condensate fraction 
(at all temperatures). On the other hand, 2D bosons with $1/r$ interactions display a finite value of $n_0$ at zero temperature~\cite{strepparola}. Fig.~\ref{fig:n0} illustrates the condensate fraction of 2D bosons with $1/r^3$ interactions, which is finite and in very good agreement with that predicted by QMC simulations~\cite{ref:astrakharchik_prl07}, even at strong coupling. In the limit $\gamma \to 0$ one expects~\cite{astrakharchik_pra2007} $n_0 \approx 1 - 1/|\ln{\gamma}|$.

\section{Summary and conclusions}
\label{sect:concl}

In summary, we have presented a self-consistent semi-analytical theory of correlations in strongly interacting fluids of two-dimensional dipolar bosons. 
Treating in an effective manner high-order correlations beyond the so-called HNC/$0$ approximation~\cite{ref:chakraborty}, we find excellent agreement between our results for the pair distribution function 
and quantum Monte Carlo data~\cite{ref:astrakharchik_prl07}. Sticking to the simplest HNC/$0$ method, we also calculate the one-body density matrix, the momentum distribution function, and the condensate fraction.

Our results are also extremely useful to study two-dimensional dipolar {\it fermions}, for which quantum Monte Carlo data have recently appeared in the literature~\cite{matveeva_arxiv_2012}. 
Fermions can be easily treated within our formalism by including a statistical contribution to the effective scattering potential~\cite{davoudi_prb_2003_fermions} 
(which is crucial to enforce the Pauli principle on the on-top value of the pair distribution function for fermions with antiparallel spin). 

A straightforward generalization of our theory 
is also useful to study density instabilities in two-dimensional dipolar fermions going beyond the Singwi-Tosi-Land-Sj\"{o}lander approximation~\cite{parish}.

\acknowledgments
We are indebted to G.E. Astrakharchik for providing us with his QMC data and to B. Tanatar for useful discussions.

\end{document}